\documentclass{article} 
\usepackage{iclr2024_conference,times}

\usepackage{amsmath,amsfonts,bm}

\def\eqref#1{equation~\ref{#1}}

\def\1{\bm{1}}

\DeclareMathAlphabet{\mathsfit}{\encodingdefault}{\sfdefault}{m}{sl}
\SetMathAlphabet{\mathsfit}{bold}{\encodingdefault}{\sfdefault}{bx}{n}

\usepackage{hyperref}
\usepackage{url}
\usepackage{graphicx}

\title{Multi-Objective Generative AI for Designing Novel Brain-Targeting Small Molecules}

\author{Ayush Noori$^{\dagger}$ \& Iñaki Arango$^{\dagger}$ \\
Harvard University\\
Cambridge, MA 02138, USA \\
\texttt{\{anoori, inakiarango\}@college.harvard.edu} \\
\AND
William E. Byrd \\
Hugh Kaul Precision Medicine Institute \\
Heersink School of Medicine \\
University of Alabama at Birmingham \\
Birmingham, AL 35233, USA \\
\texttt{webyrd@uab.edu} \\
\AND
Nada Amin\thanks{Correspondence: \texttt{namin@seas.harvard.edu}; $\dagger$ These authors contributed equally to this work.\vspace{2mm}\newline This work was accepted to the Generative and Experimental Perspectives for Biomolecular Design Workshop at the 12th International Conference on Learning Representations.} \\
Harvard John A. Paulson School of Engineering and Applied Sciences \\
Cambridge, MA 02139, USA \\
\texttt{namin@seas.harvard.edu} \\
}

\iclrfinalcopy 
\begin{document}

\maketitle

\begin{abstract}
The strict selectivity of the blood-brain barrier (BBB) represents one of the most formidable challenges to successful central nervous system (CNS) drug delivery, preventing the diagnosis and treatment of CNS disorders. Computational methods to generate BBB permeable lead compounds \textit{in silico} may be valuable tools in the CNS drug design pipeline. However, in real-world applications, BBB penetration alone is insufficient; rather, after transiting the BBB, molecules must perform some desired function – such as binding to a specific target or receptor in the brain – and must also be safe and non-toxic for use in human patients. To discover small molecules that concurrently satisfy these constraints, we use multi‑objective generative AI to synthesize drug-like blood-brain-barrier permeable small molecules that also have high predicted binding affinity to a disease-relevant CNS target. Specifically, we computationally synthesize molecules with predicted bioactivity against dopamine receptor D\textsubscript{2}, the primary target for almost all clinically effective antipsychotic drugs. After training several graph neural network-based property predictors, we adapt SyntheMol \citep{swanson_generative_2024}, a recently developed Monte Carlo Tree Search-based algorithm for antibiotic design, to perform a multi‑objective guided traversal over an easily synthesizable molecular space. We design a library of 26,581 novel and diverse small molecules containing hits with high predicted BBB permeability and favorable predicted safety and toxicity profiles, and that could readily be synthesized for experimental validation in the wet lab. We also validate top scoring molecules with molecular docking simulation against the D\textsubscript{2} receptor and demonstrate predicted binding affinity on par with risperidone, a clinically prescribed D\textsubscript{2}-targeting antipsychotic. In the future, the SyntheMol-based computational approach described here may enable the discovery of novel neurotherapeutics for currently intractable disorders of the CNS.
\end{abstract}

\section{Introduction}\label{sec:intro}
In this paper, we design small molecule drug candidates in a multi-objective setting to transit the blood-brain-barrier (BBB) and target the human brain, while also satisfying other therapeutically relevant constraints. As the interface between the central nervous system (CNS) and peripheral blood circulation, the BBB is responsible for regulating CNS homeostasis, protecting the brain microenvironment, and preventing the infiltration of toxins or pathogens across all levels of the neurovascular tree \citep{daneman_bloodbrain_2015, obermeier_development_2013} (Figure \ref{fig:blood_brain_barrier}). Although the strict selectivity of the BBB is critical to protecting CNS health and integrity, it also represents the most formidable challenge to successful CNS drug delivery in disease. The BBB blocks the access of most drugs to the brain, including approximately 100\% of all large-molecule and 98\% of all small-molecule neurotherapeutics, \textit{i.e.}, those with molecular masses greater than 400-500 kDa or with low lipid solubilities \citep{pardridge_blood-brain_2005, pardridge_drug_2012}. Further, the limited small molecule drugs which penetrate the BBB only target select brain disorders; among these are depression, schizophrenia, chronic pain, and epilepsy. Meanwhile, most CNS conditions – including neurodegenerative disorders such as Alzheimer's, Parkinson's, and Huntington's diseases; multiple sclerosis; HIV-associated neurocognitive disorders; brain cancer; neurotrauma; and cerebrovascular disease – remain refractory to neurotherapeutic administration and often lack effective treatment options \citep{pardridge_blood-brain_2005}. BBB-imposed transport selectivity also prevents the uptake of radiotracers in positron emission tomography (PET) and single photon emission computed tomography (SPECT) studies of the brain, impeding the diagnosis of CNS disorders and hindering neuroimaging-based disease staging.

Few strategies to evade or circumvent BBB regulation exist. Current approaches include intracerebroventricular injection, intranasal administration, exosome or nanoparticle-based delivery systems, and focused ultrasound (fUS) with intravenous microbubble agents to cause localized BBB disruption \citep{he_towards_2018}. However, these methods are frequently invasive with poor safety and efficacy profiles and remain an area of active research. For example, nasal epithelial barriers and the arachnoid membrane – which separates the nasal submucosa and olfactory cerebrospinal fluid (CSF) – limit brain uptake of drugs administered intranasally \citep{merkus_direct_2003}, while fUS-mediated BBB disruption may cause microhemorrhages, direct vascular rupture, ischemia, edema, thermocoagulative necrosis, or injury by microbubble-induced mechanical forces \citep{meng_safety_2019}. Thus, the most reliable solution to drug targeting in the brain is to engineer drug candidates which can cross the BBB without secondary BBB disruption or evasion. To that end, tools to discover or design CNS therapeutics that are BBB permeable with desirable pharmacokinetic profiles would be valuable assets in the CNS drug discovery process.

Wet lab methods to produce BBB permeable drugs such as high-throughput drug screening in BBB models have encountered practical challenges and achieved limited success. As a basic model of the BBB, static monolayers of primary endothelial cells or immortalized human endothelial cell lines grown in cell culture inserts can be used to study signaling pathways or transporter kinetics and may also be co-cultured with other BBB-associated cell types such as astrocytes and pericytes \citep{di_vitro_2014, bagchi_-vitro_2019}. However, such models fail to accurately replicate cell-to-cell signaling at the BBB interface and do not account for shear stress generated by physiological blood flow which regulates TJ formation and barrier function. In response, iPSC-based, microfluidic, organ-on-chip, and multicellular organoid models of the BBB continue to be developed \citep{bagchi_-vitro_2019}. Drug screening via these dynamic systems remains expensive and has yet to achieve widespread acceptance as an effective assay of BBB integrity. Further, all existing \textit{in vitro} and \textit{in vivo} models (\textit{e.g.}, the rodent brain) lack sufficient throughput required of modern, automated drug discovery programs. That is, wet lab approaches cannot be used to characterize BBB penetration at scale across the vast, unexplored regions of the neurotherapeutic molecular space.

By contrast, \textit{in silico} generation of BBB permeable drug candidates would allow for the rapid, economical, and high-throughput screening of CNS drug candidates, and may enable the discovery of novel neurotherapeutics for currently intractable CNS disorders. However, \textit{in silico} methods to generate BBB-penetrant molecules alone are insufficient. In real-world applications, after transiting the BBB, molecules must perform some desired function, such as binding to a specific target or receptor in the brain. Molecules must also be drug-like: they must be safe and non-toxic for use in human patients. To discover small molecules that concurrently satisfy these constraints, we use multi-objective generative AI to synthesize blood-brain-barrier permeable small molecules that also feature other therapeutically relevant properties, including safety, toxicity, and binding affinity to specific targets in the CNS, such as the dopamine receptor D\textsubscript{2}. 

\section{Methodology}\label{sec:methods}
Due to the intricate dynamics of transport regulation across the BBB, the complexity of drug safety and toxicity in the human body, and the large space of potential CNS neurotherapeutics, the generation of BBB permeable drug-like molecules is a task well-suited to machine learning (ML) \citep{butler_machine_2018}. By training on small molecules with known properties, supervised ML algorithms attempt to learn inductive biases about the underlying biochemical rules which govern BBB penetration capacity, which can then be leveraged to guide a molecular generation process and propose candidate molecules that satisfy bioactivity and pharmacokinetic constraints. Here, we sought to design molecules that meet three such objectives:

\begin{enumerate}
    \item Molecules must be able to transit the BBB.
    \item Molecules must exhibit binding affinity to a specific target in the brain. In this work, we select for binding affinity against the dopamine receptor D\textsubscript{2} (D2R), however, we emphasize that our approach is modular. Should a different target be of interest, our method can be adapted to generate small molecules with bioactivity against this new target.
    \item Molecules must be safe and non-toxic.
\end{enumerate}

For each objective, we obtained labeled data to act as a supervisory signal for the training of an ML model, which we refer to as a ``property predictor.'' Briefly, for BBB permeability, we used the Blood-Brain-Barrier Dataset (B3DB), a curated resource of 4,956 BBB permeable (BBB+) and 2,851 BBB non-permeable small molecules \citep{meng_curated_2021}. For D2R binding affinity, we retrieved 8,034 D2R-active molecules and sampled 12,000 D2R-inactive molecules from the ExCAPE-DB database \citep{sun_excape-db_2017}. Finally, for safety and toxicity, we retrieved a range of absorption, distribution, metabolism, excretion, and toxicity (ADME-Tox) datasets from the Therapeutics Data Commons (TDC) \citep{huang_therapeutics_2021, huang_artificial_2022}.

Next, we trained graph neural networks on these datasets (see Appendix \ref{sec:suppmethods}) to predict the properties of interest. Specifically, for each model, we used a graph convolutional network encoder and concatenated with structured molecular features to produce likelihood scores of BBB permeability or D2R bioactivity (Figures \ref{fig:synthemol_results}A, \ref{fig:model_architecture}). To evaluate out-of-sample model performance in a data-efficient manner, property predictor models were trained with 10-fold cross-validation, and the average prediction across all 10 folds was taken as the final likelihood score. For the ADME-Tox property predictor, we use ADMET-AI, an ML platform for ADME-Tox prediction trained on the TDC datasets \citep{swanson_admet-ai_2023}.

Finally, we combined the outputs of all three property predictors to guide a search across the small molecule space. As described in Appendix \ref{sec:suppmethods}, we adapted SyntheMol, a recently developed Monte Carlo Tree Search (MCTS)-based algorithm for antibiotic design, to perform a multi-objective guided traversal over 29.6 billion molecules in the Enamine REadily AccessibLe (REAL) space \citep{swanson_generative_2024, grygorenko_generating_2020}.  Like in SyntheMol, in each iteration of the algorithm, we perform an MCTS rollout to construct a molecule using the 132,479 commercially available molecular building blocks and 13 chemical synthesis reactions; however, we then evaluate the molecule using both the BBB permeability and D2R bioactivity property predictors. Using this search strategy, we generated a molecular library of 26,581 small molecule drug candidates with putative BBB permeability, D2R bioactivity, safety, and non-toxicity, which we characterized to evaluate the success of our method.

\section{Results and Discussion}\label{sec:results}
To predict BBB permeability, D2R bioactivity, and ADME-Tox characteristics, three separate property predictors were used, trained on the B3DB, pre-processed ExCAPE-DB, and TDC databases, respectively \citep{meng_curated_2021, sun_excape-db_2017, huang_therapeutics_2021, swanson_admet-ai_2023}. Encouragingly, our BBB and D2R property predictor models demonstrated strong out-of-sample performance, as visualized by AUROC curves in Figure \ref{fig:property_predictors}. The BBB predictor achieved greater than 0.95 AUC across all folds, while the D2R bioactivity predictor achieved near-perfect performance of 1.00 AUC, again across all folds. We highlight the performance of the first cross-validation fold for the BBB and D2R predictors (Figure \ref{fig:property_predictors}A and \ref{fig:property_predictors}B) and also show the performance across a set of randomly sampled folds for both models (Figures \ref{fig:property_predictors}C, \ref{fig:property_predictors}D). These results suggest that our trained models are likely useful oracles for predicting both BBB permeability and binding affinity to D2R.

\begin{figure}[h]
    \centering
    \includegraphics[width=\linewidth]{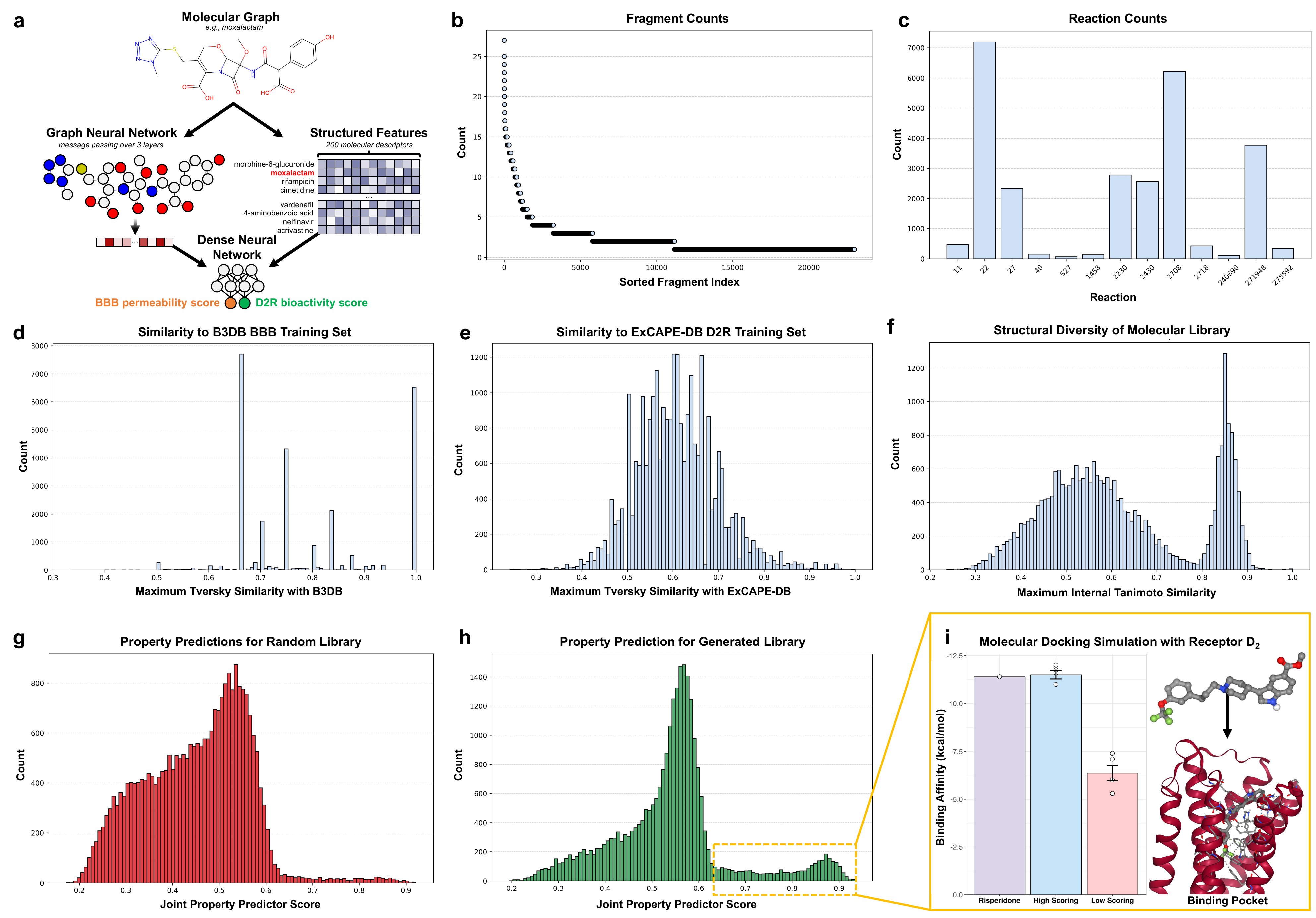}
    \caption{\textbf{Results of SyntheMol-based guided Monte Carlo Tree Search for multi-objective small molecule design.} Guided by predictors of BBB permeability, D2R binding, and toxicity, MCTS was performed over 29.6 billion compounds to generate a library of 26,581 molecules.}
    \label{fig:synthemol_results}
\end{figure}

Guided by these high-performing property predictors, we performed a SyntheMol-based MCTS search over 29.6 billion molecules in the pre-processed subset of the Enamine REAL combinatorial molecular space to generate a molecular library of 26,581 small molecules over 20,000 MCTS rollouts \citep{swanson_generative_2024, grygorenko_generating_2020}. First, we investigated diversity in building blocks and reactions sampled by the SyntheMol-based search. Across all 26,581 molecules, no individual molecular fragment was employed more than 27 times; that is, at most, any individual building block will be contained in only 0.102\% of the molecular library (Figure \ref{fig:synthemol_results}B). Further, all 13 chemical reactions are employed in the search process, and 11 of the 13 chemical reactions are used to generate more than 300 molecules (Figure \ref{fig:synthemol_results}C).

Next, we explored the similarity of the generated molecules to the labeled molecules in the B3DB (Figure \ref{fig:synthemol_results}D) and ExCAPE-DB training (Figure \ref{fig:synthemol_results}E) data, for the BBB permeability and D2R bioactivity prediction tasks, respectively. For each molecule in the library, we computed the maximum Tversky similarity (an asymmetric similarity measure) between the Morgan fingerprints of that molecule and the fingerprints of all other molecules in each of the training sets. We consider a Tversky similarity score greater than 0.9 to be undesirable, and find that the 73.61\% and 99.47\% of the molecules have similarity scores less than 0.9 against both the B3DB (Figure \ref{fig:synthemol_results}D) and ExCAPE-DB (Figure \ref{fig:synthemol_results}E) datasets, respectively. The diversity of generated molecules was more preferable for ExCAPE-DB as compared to B3DB, which was expected due to the increased size and diversity of the ExCAPE-DB database as compared to B3DB. These results indicate that the SyntheMol-based search did not merely memorize and reproduce molecules from the training data.

We also characterized the structural diversity within the 26,581 compounds that comprise our molecular library (Figure \ref{fig:synthemol_results}F). For each molecule in the library, we computed the maximum Tanimoto distance (a symmetric distance metric) between the Morgan fingerprints of that molecule and the fingerprints of all other molecules in the library. We find that 73.61\% of the molecules have an internal similarity less than 0.8, and 99.29\% of the molecules have an internal similarity less than 0.9. These results suggest that the molecular library generated here is diverse and amenable to experimental screening; that is, the SyntheMol-based search did not produce a structurally homogeneous library. Our results also align with those of \citet{swanson_generative_2024}, even though the prior work focused on antibiotic discovery for infectious diseases and bacterial pathogens, while our synthesis efforts concern a wholly different biological domain of neurotherapeutics for CNS disorders.

Finally, we use the trained BBB, D2R, and ADME-Tox property predictors to evaluate each of the molecules in the library and produce an aggregate likelihood score $\alpha$ of each biomolecular property (Figure \ref{fig:synthemol_results}H). We visualize the average joint likelihood and compare against an unguided MCTS baseline where building blocks were randomly sampled (Figure \ref{fig:synthemol_results}G). Not all molecules are predicted as BBB permeable, active against D2R, and non-toxic – indeed, there is a peak at $\alpha = 0.55$, suggesting that many molecules are predicted to have some, but not all, properties of interest. However, in the guided search, there is a second peak at $\alpha = 0.87$: 2904 molecules have $\alpha > 0.65$ and 1662 molecules have $\alpha > 0.80$. By contrast, in the unguided random baseline, only 573 molecules have $\alpha > 0.65$ and only 224 molecules have $\alpha > 0.80$. Thus, our multi-objective SyntheMol-based guided Monte Carlo Tree Search synthesized high-scoring molecules at a 5.07-fold and 7.42-fold higher hit rate for $\alpha > 0.65$ and $\alpha > 0.80$, respectively, as compared to an unguided baseline.

We further validated top scoring generated molecules with protein-ligand molecular docking simulation using AutoDock Vina \citep{trott_autodock_2010, eberhardt_autodock_2021, murail_seamdock_2021}. We retrieved the crystal structure of the dopamine receptor D\textsubscript{2} (Protein Data Bank (PDB) ID: \texttt{6cm4}) \citep{wang_structure_2018}, and simulated the binding of D2R against the top 5 generated molecules ranked by D2R property predictor score, as well as the bottom 5 scoring molecules and risperidone (a second generation antipsychotic prescribed for the treatment of bipolar disorder and schizophrenia which acts by binding to D2R) as negative and positive controls, respectively. After docking simulation, the predicted binding affinity of risperidone was -11.40 kcal/mol, comparable to the mean predicted affinities of the top scoring molecules at -11.50 kcal/mol (Figure \ref{fig:synthemol_results}I, error bars show mean $\pm$ SE). By contrast, under simulation, the bottom scoring molecules bound more weakly, with mean predicted affinities of -6.36 kcal/mol. These results indicate that our SyntheMol-based molecular design strategy discovered small molecules with strong predicted binding affinity to the dopamine receptor D\textsubscript{2} as well as high predicted BBB permeability and non-toxicity. These compounds may represent early drug candidates for challenging CNS conditions such as schizophrenia, Parkinson's disease, and substance abuse disorders, and could merit further investigation and experimental characterization in the wet lab.

\subsection{Future directions}

Various future directions exist to extend this work, address the limitations of this proof-of-concept study, and validate our findings. First, when training property predictors, rather than a uniform, random split, a scaffold split could be used to partition molecules based on their Bemis-Murcko scaffolds \citep{wu_moleculenet_2018, landrum_rdkit_2022}. Such two-dimensional scaffolds simplify the chemical structure of a compound by ignoring side chains, atom type, hybridization, and bond order; only ring systems are preserved \citep{bemis_properties_1996}. By ensuring greater structural diversity among the training and test sets, scaffold splitting would improve the out-of-distribution generalization capacity of our trained property predictors and likely increase the quality of our generated molecular library. Other areas of improvement include evaluation of different neural architectures in the encoder of our property predictor models (\textit{e.g.}, graph transformers), addition of DeepPurpose-based ADME-Tox predictors over ADMET-AI \citep{huang_deeppurpose_2021}, provisioning of additional computational resources to enable more MCTS rollouts and and a larger generated library, logic-based integration of the individual property predictor outputs in $P(N)$ rather than taking the average score, and parallelization of the MCTS search to improve performance and increase coverage of the MCTS rollouts over the Enamine REAL space (see Appendix \ref{sec:suppmethods}). In particular, the use of Transformer-based encoders would allow us to identify functional groups of synthesized molecules with the highest attention weights that may be most relevant to classification of BBB permeability and could inform future rational drug design efforts in medicinal chemistry or suggest novel mechanisms for BBB transport and uptake into the brain.

Nonetheless, these results establish groundwork in multi-objective \textit{in silico} discovery of novel CNS therapeutic candidates. By training multiple property predictors to drive a search over the Enamine REAL space, we jointly optimized for several important drug prerequisites – namely, BBB permeability, toxicity, synthesizability, and bioactivity against the important dopamine receptor D\textsubscript{2}. Further, we note that the method described here could be used to discover small molecules with binding affinity to targets other than D2R. Although D2R is indeed a disease-relevant target of high priority, we emphasize that our approach is modular and not restricted to D2R or the conditions in which D2R is implicated. Rather, by leveraging the 70,850,163 bioactivity annotations in the chemogenomics database ExCAPE-DB, we could feasibly train predictors of bioactivity against 1666 additional protein targets beyond D2R \citep{sun_excape-db_2017}. Coupled with SyntheMol-based guided search, this strategy would allow us to develop molecular libraries relevant to a range of other conditions that could again be validated by molecular docking simulation. The approach advanced here could be used to profile unexplored regions of the CNS neurotherapeutic space and suggest new drug and radiotracer candidates to better diagnose and treat CNS disease. Finally, the most exciting and urgent future direction of this work is experimental validation in the wet lab. We are eager to partner with experimentalists to evaluate our molecular library in both \textit{in vitro} and \textit{in vivo} models of schizophrenia and Parkinson's disease, in an effort to work towards pharmacological interventions for these currently intractable neurological disorders.

\subsubsection*{Code Availability}
Code is available via GitHub at \href{https://github.com/ayushnoori/molecule-synthesis}{https://github.com/ayushnoori/molecule-synthesis}.

\subsubsection*{Acknowledgments}
We gratefully acknowledge Kyle Swanson and James Zou for their support and collaboration, allowing us to extend their prior work. K.S. and J.Z. shared an unpublished copy of their manuscript with us, ``Generative AI for designing and validating easily synthesizable and structurally novel antibiotics''; this work is now published at the Conference on Neural Information Processing Systems (NeurIPS) 2023 Generative AI and Biology Workshop and in \textit{Nature Machine Intelligence}. K.S. and J.Z. also provided us with early access to their code base, now available at \href{https://github.com/swansonk14/SyntheMol}{https://github.com/swansonk14/SyntheMol} on GitHub. We thank K.S. and J.Z. for their critical review of our manuscript and for their constructive feedback. We also thank Venkatesh Murthy and Tyler Holloway for valuable discussions.

Figure \ref{fig:blood_brain_barrier} was created with BioRender.com. 

\bibliography{references}
\bibliographystyle{iclr2024_conference}

\newpage
\appendix

\section{Blood-Brain-Barrier}\label{sec:suppintro}
The blood-brain-barrier (BBB) is comprised of specialized endothelial cells and mural cells which form the continuous nonfenestrated vasculature of the brain and spinal cord. Unlike non-CNS endothelium, various barrier functions of the cerebral capillary endothelial cells allow selective transcellular but not paracellular transport across the BBB, which allows close regulation of molecular traffic entering the brain \citep{abbott_astrocyte-endothelial_2006}. These functions include:

\begin{enumerate}
    \item Physical barriers, such as continuous intercellular tight junctions (TJs) formed of occludin, claudins, tricellulins, and junctional adhesion molecules; other junctional complexes, such as adherens and gap junctions; the lack of membrane fenestra, or pores; and low rates of transcellular vesicle trafficking and other mechanisms of transcytosis (Figure \ref{fig:blood_brain_barrier}) \citep{abbott_structure_2010, langen_development_2019, stamatovic_junctional_2016}. 
    \item Molecular barriers, such efflux transporters to counteract the diffusion of lipophilic substances (\textit{e.g.}, drugs or xenobiotics; see Figure \ref{fig:blood_brain_barrier}) and low expression of leukocyte adhesion molecules (LAMs) to hinder leukocyte extravasation into the CNS parenchyma \citep{daneman_bloodbrain_2015}.
    \item Metabolic barriers, including intracellular (\textit{e.g.}, monoamine oxidase and cytochrome P450) and extracellular (\textit{e.g.}, peptidases and nucleotidases) drug-metabolizing enzymes \citep{abbott_astrocyte-endothelial_2006, minn_drug_1991}.
    \item Extravascular barriers formed by other cell types of the neurovascular unit (NVU), such as astrocytes and pericytes – which are embedded in the basement membrane of the abluminal surface of the BBB endothelium \citep{obermeier_development_2013}. For example, extravascular structures such as the endothelial glycocalyx and astrocyte endfeet (Figure \ref{fig:blood_brain_barrier}) may inhibit the delivery of large molecules across the BBB \citep{kutuzov_contributions_2018}. Further, ECM receptors in the basement membrane, including dystroglycans and integrins, mediate cell-matrix adhesion and influence barrier function (\textit{e.g.}, by regulating TJ organization) \citep{langen_development_2019}.
\end{enumerate}

These barrier systems prevent the passive (\textit{i.e.}, transendothelial) diffusion of all but small gaseous molecules (\textit{e.g.}, O\textsubscript{2} and CO\textsubscript{2}) and certain lipophilic agents (\textit{e.g.}, barbiturates, ethanol) across the BBB (Figure \ref{fig:blood_brain_barrier}) \citep{abbott_astrocyte-endothelial_2006}; all other transport is restricted to facilitated or active pathways. These pathways – which include carrier-mediated transport (\textit{i.e.}, ATP-independent solute carriers) and receptor-mediated transcytosis – provision select nutrients and metabolites to satisfy the energetic and chemical requirements of the brain \citep{campos-bedolla_role_2014}. Nutrient delivery notwithstanding, the capillaries of the BBB remain the least permeable in the human body.

\newpage
\section{Supplemental Methods}\label{sec:suppmethods}
\subsection{Training Data}

The labeled datasets used to train property predictors for each design objective are described below.

\textbf{\textit{Blood-brain-barrier permeability.}} To train a property predictor for BBB permeability, we use the Blood-Brain-Barrier Dataset (B3DB), a curated resource of 7,807 small molecules classified as either BBB permeable (BBB+) or BBB non-permeable (BBB-), with 4,956 BBB+ and 2,851 BBB- molecules included \citep{meng_curated_2021}. BBB permeability is measured the logarithm of the brain-plasma concentration ratio:
\begin{align*}
    \log{\text{BB}} &= \log{\frac{C_\text{brain}}{C_\text{blood}}}
\end{align*}

Although numerical  data is included for 1,058 of the 7,807 molecules in the dataset, we seek to produce a property predictor that can be trained across a maximally diverse set of molecules to guide the search across a similarly diverse and much larger space of possible small molecules. Therefore, we train a supervised classifier to predict the binary BBB+ or BBB- label, where $\log{\text{BB}} > 0 =$ BBB+ and $\log{\text{BB}} < 0 =$ BBB-. To represent each molecule, we use the simplified molecular-input line-entry system, or SMILES, a specification for describing the structure of chemical species using short ASCII strings (Figure \ref{fig:model_architecture}A) \citep{weininger_smiles_1988}. The final dataset included the SMILES representation of each molecule and binary classification label (\textit{i.e.}, 0 for BBB- and 1 for BBB+).

\textbf{\textit{Bioactivity against target of interest.}} To discover molecules that exhibit binding affinity to a specific target in the brain, we use ExCAPE-DB, a large-scale database of chemogenomics data that includes 70,850,163 bioactivity annotations of 998,131 unique chemical compounds against 1667 protein targets \citep{sun_excape-db_2017}. ExCAPE-DB was assembled from the PubChem \citep{kim_pubchem_2023} and ChEMBL \citep{gaulton_chembl_2012, bento_chembl_2014} databases, and is a valuable resource for quantitative structure-activity relationship (QSAR) modeling. Of note, ExCAPE-DB is already filtered for drug-like compounds; therefore, all molecules in the database are organic (\textit{i.e.}, lack metal atoms), have molecular weights less than 1,000 Da, and have more than 12 heavy atoms. We download Version 2.0 of ExCAPE‑DB from the Zenodo repository at DOI: \href{https://doi.org/10.5281/zenodo.2543724}{10.5281/zenodo.2543724}.


In this work, we select the dopamine receptor D\textsubscript{2} (D2R), encoded by the \textit{DRD2} gene in humans, as our target of interest. D2R is one of a family of five G-protein-coupled receptors that facilitate the function of the catecholamine neurotransmitter dopamine; D2R does so by coupling to the inhibitory G protein G\textsubscript{i/o}, likely to regulate presynaptic firing rate \citep{zhuang_structural_2021, beaulieu_physiology_2011}. D2R is the primary target for almost all clinically effective antipsychotic drugs and its function can be affected by binding of both agonist or antagonist ligands, which activate or inhibit D2R, respectively. Importantly, D2R signaling has been implicated in various neurological and neuropsychiatric diseases, including schizophrenia, tardive dyskinesia, attention-deficit hyperactivity disorder, Tourette's syndrome, Parkinson's disease, and substance abuse disorders \citep{beaulieu_physiology_2011, kostrzewa_dopamine_2018, di_chiara_dopamine_2004, volkow_brain_2015, seeman_targeting_2006}. Further, D2R is expressed across the brain, with highest expression in the striatum, nucleus accumbens, and olfactory tubercle, but also in the substantia nigra, ventral tegmental area, hypothalamus, cortical regions, septum, amygdala, and hippocampus \citep{beaulieu_physiology_2011}. Therefore, the discovery of novel small molecules that can bind to and modulate the function of D2R may be important tools in the clinical arsenal to restore homeostatic dopaminergic signaling in challenging disorders of the CNS.

In line with previous work \citep{olivecrona_molecular_2017, yang_improving_2020}, the ExCAPE-DB database was subset for molecules with bioactivity data against D2R. After filtering for duplicate molecules by canonical SMILES structure, 8,034 active molecules and 342,204 inactive molecules remained. Bioactivity is often measured by $\text{pIC}_{50}$, where $\text{IC}_{50}$ is a measure of the concentration of a chemical species required to inhibit a biological target \textit{in vitro}, and $\text{pIC}_{50} = -\log_{10}(\text{IC}_{50})$. In ExCAPE-DB, compounds were classified as active against a specific target if $\text{pIC}_{50} > 5$ and inactive if $\text{pIC}_{50} < 5$. However, numerical  data was only available for 293 of the 342,204 inactive compounds; therefore, we train a supervised classifier to predict the binary active or inactive label. We randomly sample 12,000 compounds from the 342,204 inactive compounds, to produce a final dataset of approximately 20,000 molecules with labeled bioactivity against D2R. Like before, the final dataset included the SMILES representation of each molecule and binary classification label (\textit{i.e.}, 0 for inactive and 1 for active).

\textbf{\textit{Safety and toxicity.}} To predict safety and toxicity, we retrieve data from the Therapeutics Data Commons, a centralized resource of curated datasets and benchmarks for the development of AI models for therapeutics tasks \citep{huang_therapeutics_2021, huang_artificial_2022}. We retrieve a range of absorption, distribution, metabolism, and excretion (ADME) datasets as well as several toxicity datasets. ADME-related benchmarks include including data on human intestinal absorption and permeability \citep{wang_adme_2016, siramshetty_validating_2021, hou_adme_2007}, absorption-related P-glycoprotein inhibition \citep{broccatelli_novel_2011}, bioavailability \citep{ma_prediction_2008}, lipophilicity \citep{wu_moleculenet_2018}, solubility \citep{sorkun_aqsoldb_2019}, hydration free energy \citep{wu_moleculenet_2018, mobley_freesolv_2014}, plasma protein binding rate, volume of distribution at steady state \citep{lombardo_silico_2016}, inhibition of various metabolism-related cytochrome P450 genes \citep{veith_comprehensive_2009, carbon-mangels_selecting_2011, cheng_admetsar_2012}, half-life \citep{obach_trend_2008}, and clearance \citep{di_mechanistic_2012}. Toxicity-related benchmarks include data on acute toxicity \citep{richard_toxcast_2016, zhu_quantitative_2009}, blockade of the human ether-à-go-go related gene (hERG) channel responsible for coordinating cardiovascular activity \citep{wang_admet_2016, du_hergcentral_2011, karim_cardiotox_2021}, mutagenicity \citep{xu_silico_2012}, drug-induced liver injury \citep{xu_deep_2015}, chemically-induced skin reaction \citep{alves_predicting_2015}, carcinogenicty \citep{cheng_admetsar_2012, lagunin_computer-aided_2009}, and toxicity-related clinical trial failures \citep{gayvert_data-driven_2016}. All datasets were retrieved in a harmonized format from the Therapeutics Data Commons.

\subsection{Property Predictors}

\textbf{\textit{Feature engineering.}} To guide search over the molecular space, computational prediction of molecular properties was facilitated by encoding the chemical space into mathematical descriptors. First, we used RDKit, an open-source cheminformatics toolkit, to compute a global pharmacophore descriptor represented as 200-element vector from the SMILES representations of each molecule (Figure \ref{fig:model_architecture}A) \citep{landrum_rdkit_2022}. Next, the SMILES representations of each molecule in the training data were converted to molecular graphs, where nodes correspond to atoms and edges correspond to chemical bonds. Based on the molecular graphs, we then created molecular fingerprints, or vector representations of the structural properties of each molecule which encode the local chemical environment of a molecule by iteratively applying a hashing function to molecular substructures \citep{cereto-massague_molecular_2015}. In particular, we use the Morgan fingerprint, also known as the extended-connectivity fingerprint ECFP4, a 1024-bit vector where bits are assigned based on the presence of circular substructures around each atom in a molecule \citep{morgan_generation_1965}. We use a radius of 2; therefore, the hashing function is applied to all substructures within 2 bonds of each atom in the molecule.

\textbf{\textit{Graph neural networks.}} To predict blood-brain-barrier permeability and D2R bioactivity, two separate property predictors were trained (Figure \ref{fig:model_architecture}A). As in \citet{swanson_generative_2024}, using the Chemprop framework, we developed graph neural networks (GNNs) with three layers of message-passing, followed by a two-layer feed-forward neural network that combines the output of the GCN with the 200-element RDKit chemical descriptor to predict the likelihood of either blood-brain-barrier permeability or D2R bioactivity. We included the RDKit chemical descriptors based on our observation that structured features like molecular weight are correlated with blood-brain-barrier permeability, as measured by correlation with numerical  ratio for 1,058 of the 7,807 molecules in the B3DB dataset (Figure \ref{fig:model_architecture}B and \ref{fig:model_architecture}C). To evaluate out-of-sample model performance while maximizing usage of the available training data, models were trained with 10-fold cross-validation with 30 epochs of training per fold and 80\%-10\%-10\% split across training, validation, and test sets, respectively. The average prediction across all 10 folds was taken as the final likelihood score, and performance was visualized in area under the receiver operating characteristic (AUROC) curves (Figure \ref{fig:property_predictors}).

Next, to predict absorption, distribution, metabolism, excretion, and toxicity, we evaluated two approaches. First, we used the DeepPurpose library to train our own individual property predictors for each of the ADME and toxicity (ADME-Tox) datasets described in the ``Training Data'' section. DeepPurpose is a library for drug-target interaction prediction that supports 15 compound and protein encoders and over 50 neural architectures \citep{huang_deeppurpose_2021}. We trained 2-layer message passing GNNs on the molecular graph of each compound to predict ADME-Tox features. Models were trained for 30 epochs, with a learning rate of  and a batch size of 128. After training, an aggregate model was assembled to take as input a single molecular query and output the results of the ADME-Tox models in a single concatenated vector. We also considered ADMET-AI, a recently developed ML platform for ADME-Tox property prediction trained on the TDC datasets described above \citep{swanson_admet-ai_2023}. For each molecule, computed the weighted average of the ADMET-AI prediction scores (with a weight of 1 for non-toxic properties and -1 for toxic properties), and linearly scaled the scores from 0 to 1. In this work, we use ADMET-AI as the ADME-Tox property prediction oracle; the inclusion of our own DeepPurpose-based ADME-Tox predictors represents an interesting avenue for future work.

\subsection{SyntheMol-based Monte Carlo Tree Search}

After training the property predictors, we adapt SyntheMol, a recently developed Monte Carlo Tree Search (MCTS)-based algorithm for antibiotic design, to perform a multi-objective guided traversal over an easily synthesizable molecular space \citep{swanson_generative_2023, swanson_generative_2024}. SyntheMol was designed to overcome a key challenge in generative AI for molecular design: most generative models for \textit{de novo} molecular design often produce compounds that are intractable to synthesize in the wet lab, and thus cannot be validated experimentally \citep{gao_synthesizability_2020}. To ensure synthesizability, SyntheMol assembles compounds using a pre-processed subset of the Enamine REadily AccessibLe (REAL) Space, composed of 132,479 commercially available molecular building blocks with known reactivities and 13 chemical synthesis reactions \citep{grygorenko_generating_2020}. The resulting combinatorial chemical search space spans 29.6 billion molecules that are tractable to synthesize experimentally.

In this work, we adapt the SyntheMol algorithm in \citet{swanson_generative_2024} to perform multi-objective molecular search guided by the property prediction models for BBB permeability, D2R bioactivity, and ADME-Tox described above. Like in SyntheMol, in each iteration of the algorithm, we perform an MCTS rollout to construct a molecule using the 132,479 building blocks and 13 chemical reactions; however, we then evaluate the molecule using \textit{both} the BBB permeability and D2R bioactivity property predictors.

In the MCTS search, molecular building blocks and their possible combinations via chemical synthesis reactions are represented by nodes in a synthesis tree. During the MCTS rollouts, each potential child node $N$ is evaluated by a scoring function that balances both exploitation of nodes that lead to high scoring molecules as well as exploration of nodes that have rarely been visited during the search. We use the SyntheMol scoring function as described in \citet{swanson_generative_2024}; however, we adapt the molecular property score $P(N)$ to perform guided molecular search in a multi-objective setting by concurrently using the property predictors for BBB permeability, D2R bioactivity, and ADME-Tox. Let $M_\text{BBB}$ be the predictor of BBB permeability, let $M_\text{D2R}$ be the predictor of D2R bioactivity, and let $M_\text{Tox}$ be the ADME-Tox predictor. Then, $P(N)$ is given by:
\begin{align*}
    P(N) = \frac{1}{|N_\text{mols}|} \sum_{i = 1}^{|N_\text{mols}|}{\frac{M_\text{BBB}(N^i_\text{mols}) + M_\text{D2R}(N^i_\text{mols}) + M_\text{Tox}(N^i_\text{mols})}{3}}
\end{align*}

Here, $N^i_\text{mols}$ is the $i$-th molecule in node $N$ (in case reactions produce more than 1 molecule). This approach is also inspired by MolSearch, a multi-objective MCTS-based method for molecular generation with a two-stage design which alternates between optimizing for biological properties and optimizing for non-biological properties, instead of taking the average across the property predictors as we do here \citep{sun_molsearch_2022}.

We use this SyntheMol-based guided search strategy to produce a molecular library of 26,581 novel and diverse molecules. To characterize the library, we visualize the distribution of molecular building blocks and chemical synthesis pathways employed. We also investigate the molecular diversity of the library using maximum internal Tanimoto similarity as well as maximum Tversky similarity with the training data. Finally, we evaluate each of the molecules using both the BBB and D2R predictors, compare the average score to an unguided random sampling baseline, and validate top scoring hits with protein-ligand molecular docking simulation against D2R.

\newpage
\section{Supplemental Figures}\label{sec:suppfigs}
\renewcommand{\thefigure}{S\arabic{figure}}
\setcounter{figure}{0}

\begin{figure}[h]
    \centering
    \includegraphics[width=\linewidth]{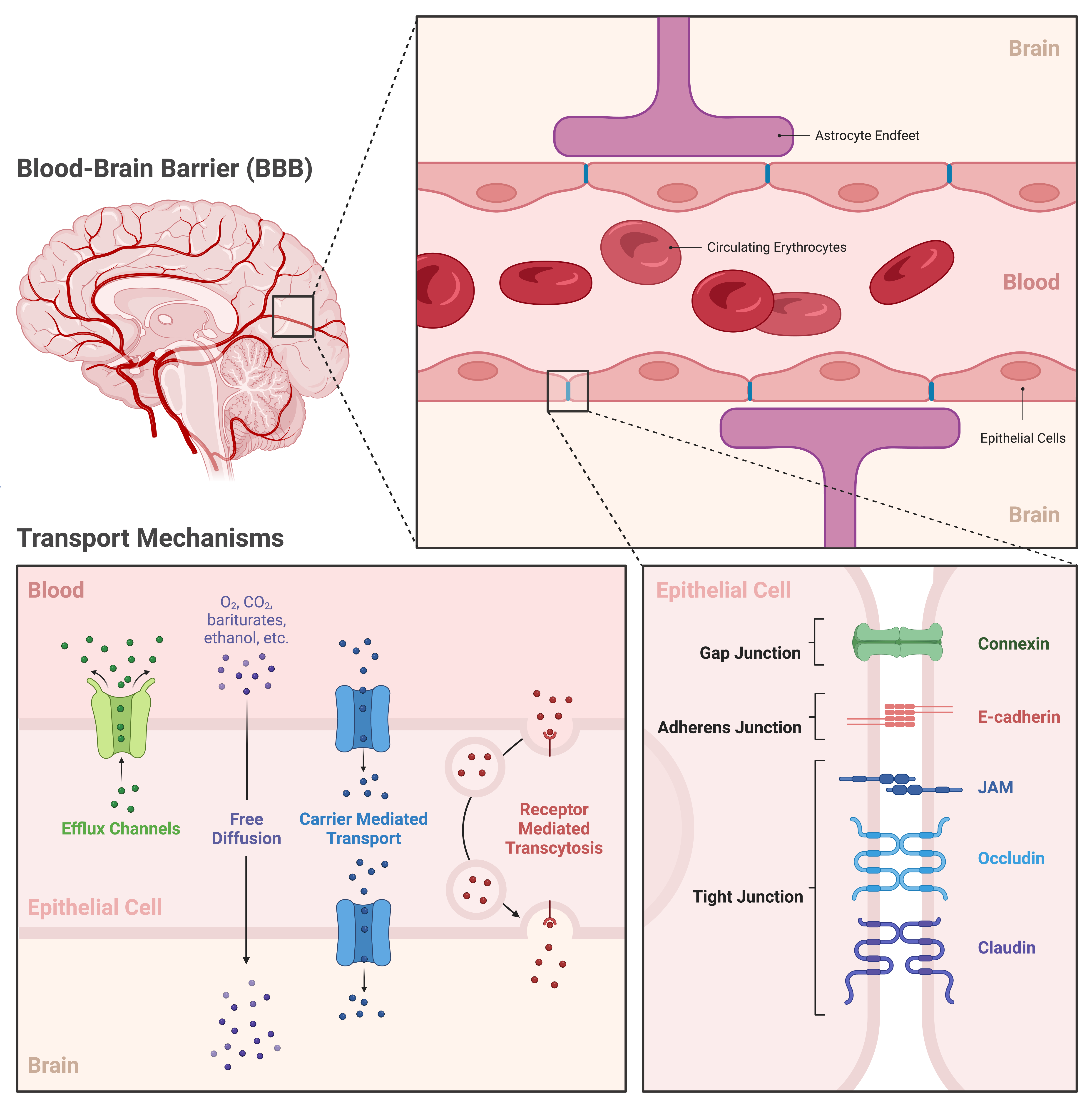}
    \caption{\textbf{Transport and barrier functions of the CNS epithelium.} Physical, molecular, metabolic, and extravascular barriers of the BBB regulate access to the CNS milieu. Created using Biorender.com.}
    \label{fig:blood_brain_barrier}
\end{figure}

\begin{figure}[t]
    \centering
    \includegraphics[width=\linewidth]{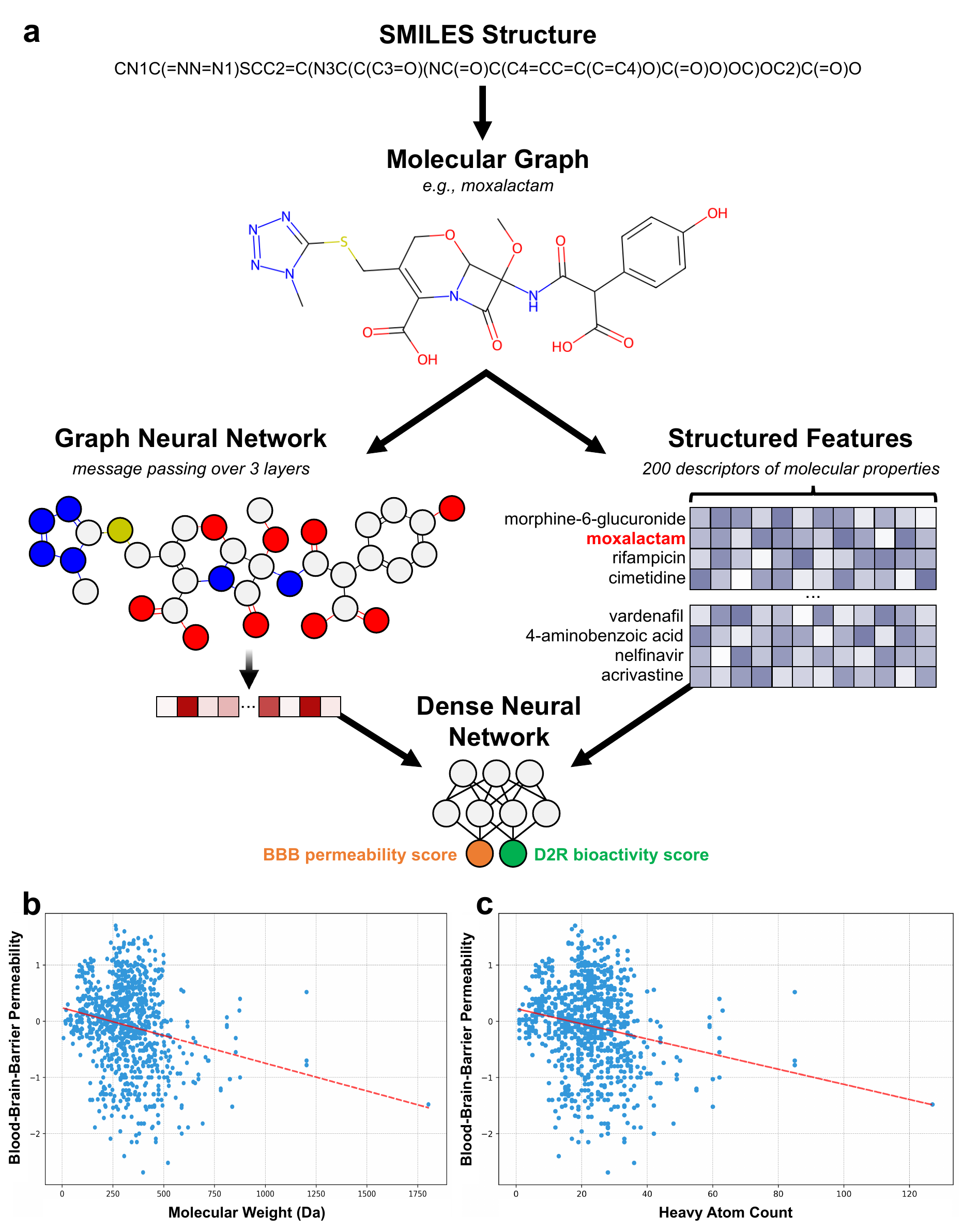}
    \caption{\textbf{Model architecture of property predictors.} To predict blood-brain-barrier permeability and D2R bioactivity, two separate property predictors were trained. \textbf{(A)} First, the SMILES representations of each molecule in the training data were converted to molecular graphs; for example, the SMILES structure and molecular graph of moxalactam, a third-generation cephalosporin antibiotic in the B3DB database, are shown here. Next, we developed graph neural networks (GNNs) with three layers of message-passing, followed by a two-layer feed-forward neural network that combined the output of the GCN with a 200-element RDKit chemical descriptor to predict the likelihood of either blood-brain-barrier permeability or D2R bioactivity. The structured chemical descriptors are included based on the observation that properties like \textbf{(B)} molecular weight and \textbf{(C)} number of heavy atoms are correlated with blood-brain-barrier permeability, as measured by correlation with the numerical blood-brain-barrier concentration ratios for 1,058 of the 7,807 molecules in the B3DB dataset for which the data was available.}
    \label{fig:model_architecture}
\end{figure}

\begin{figure}[t]
    \centering
    \includegraphics[width=\linewidth]{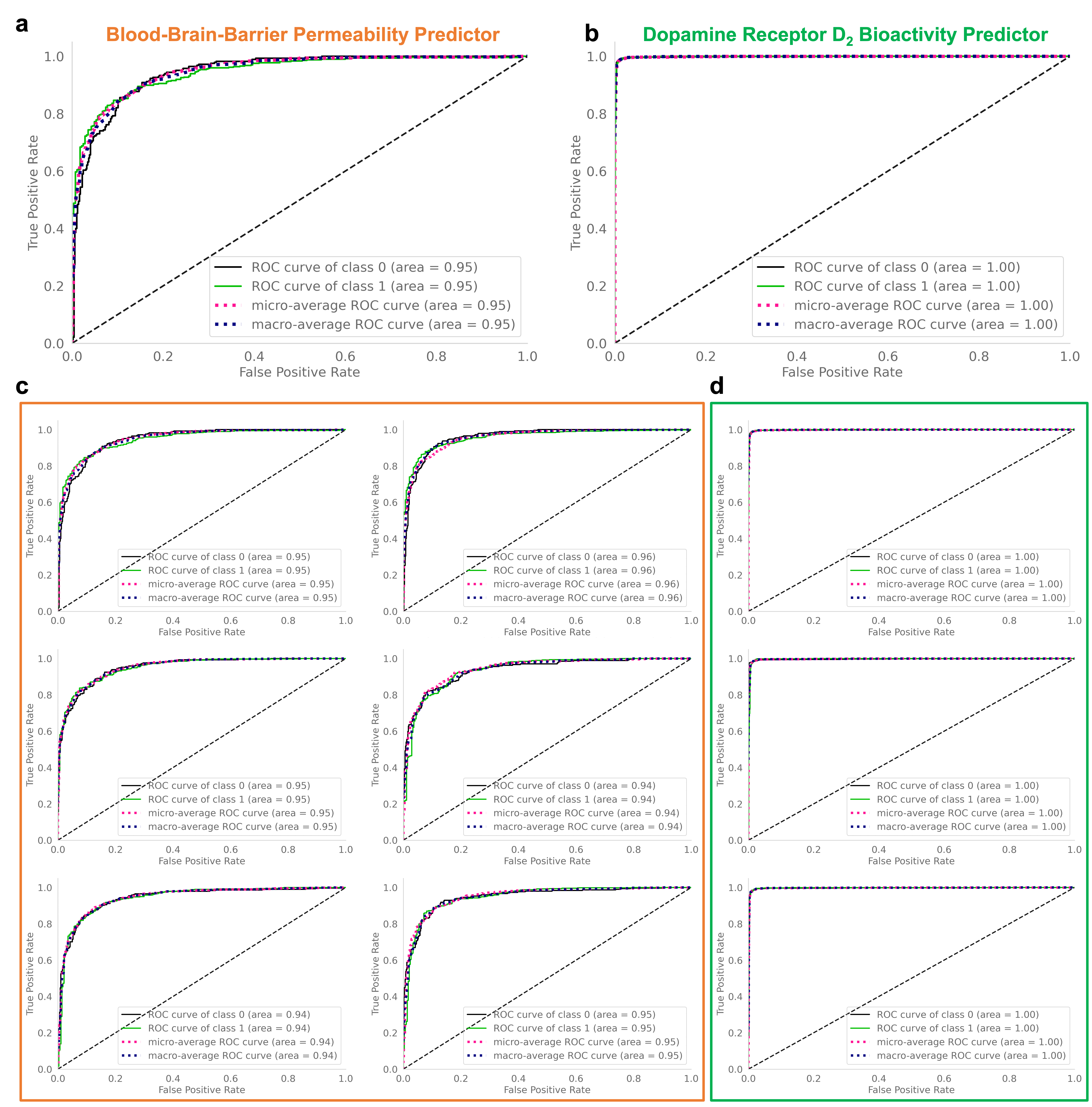}
    \caption{\textbf{Property predictor performance.} To evaluate out-of-sample model performance while maximizing usage of the available training data, property predictor models were trained with 10-fold cross-validation with 30 epochs of training per fold and 80\%-10\%-10\% split across training, validation, and test sets, respectively. The average prediction across all 10 folds was taken as the final likelihood score, and model performance on the independent test set was visualized in area under the receiver operating characteristic (AUROC) curves. Here, we highlight the performance of the first cross-validation fold for the \textbf{(A)} blood-brain-barrier permeability predictor and \textbf{(B)} the dopamine receptor D\textsubscript{2} bioactivity predictor. We also show the performance across a set of randomly sampled folds for both the \textbf{(C)} BBB and \textbf{(D)} D2R predictors.}
    \label{fig:property_predictors}
\end{figure}

\end{document}